\documentclass[prl,aps,10pt,floatfix]{revtex4-1}
\usepackage{amsmath}
\usepackage{amssymb}
\usepackage{bm}
\usepackage{graphicx, float}
\usepackage{listings}
\usepackage{color}
\usepackage{color}
\definecolor{mygreen}{rgb}{0,0.6,0}
\definecolor{mygray}{rgb}{0.5,0.5,0.5}
\definecolor{mymauve}{rgb}{0.58,0,0.82}
\usepackage{amssymb}
\usepackage[figuresright]{rotating}

\begin{document}

\title{Rotational line strengths for the CN $B \, ^2\Sigma^+ - X \, ^2\Sigma^+$ (5,4) band}
\author{James O. Hornkohl}
\address{Hornkohl Consulting, 344 Turkey Creek Road,
Tullahoma, TN 37388, USA}
\author{Christian G. Parigger \footnote{Corresponding author: Christian Parigger, 931 841-5690, cparigge@tennessee.edu}}
\address{University of Tennessee Space Institute,
Center for Laser Applications,
411 B.H. Goethert Parkway,
Tullahoma, TN 37388, USA}

\begin{abstract}
Rotational line strengths, computed from eigenvectors of Hund's case (a) matrix representations of the upper and lower Hamiltonians using Wigner-Witmer basis functions, show a larger than expected influence from the well known perturbation in the (5,4) band. Comparisons with National Solar Observatory experimental Fourier transform spectroscopy data reveal nice agreement of measured and predicted spectra. \\


\end{abstract}
\maketitle

\section{Introduction}
The CN violet $B \, ^2\Sigma^+ - X \, ^2\Sigma^+$ band system is one of the most studied band systems. Ram \textit{et al.} \cite{RDWEAB} and
Brooke \textit{et al.} \cite{JSABrookeCN} have summarized the available experimental and theoretical information.
Of the many known bands in the violet system, only the (5,4) band is considered here. This band exhibits a weak, quantitatively understood perturbation \cite{Ito5-4} caused by mixing of the $v=17$ level of $A \, ^2\Pi$ with the $v=5$ level of $B \, ^2\Sigma^+$. The particular perturbation of the CN (5,4) band is evaluated in this work by isolating the spectral features of this band that is part of the CN viloet system.
Numerical diagonalizations of upper and lower Hamiltonians with and without the perturbation are investigated and compared with available experimental spectra. The simulations rely on determining rotational strengths without parity-partitioned Hamiltonians. It is anticipated that the investigated (5,4) band modifications can be possibly confirmed with the new PGOPHER program recently released by Western \cite{Pgopher}.

\section{CN (5,4) band spectra}
\begin{figure}
\begin{center}
\includegraphics[scale=0.75]{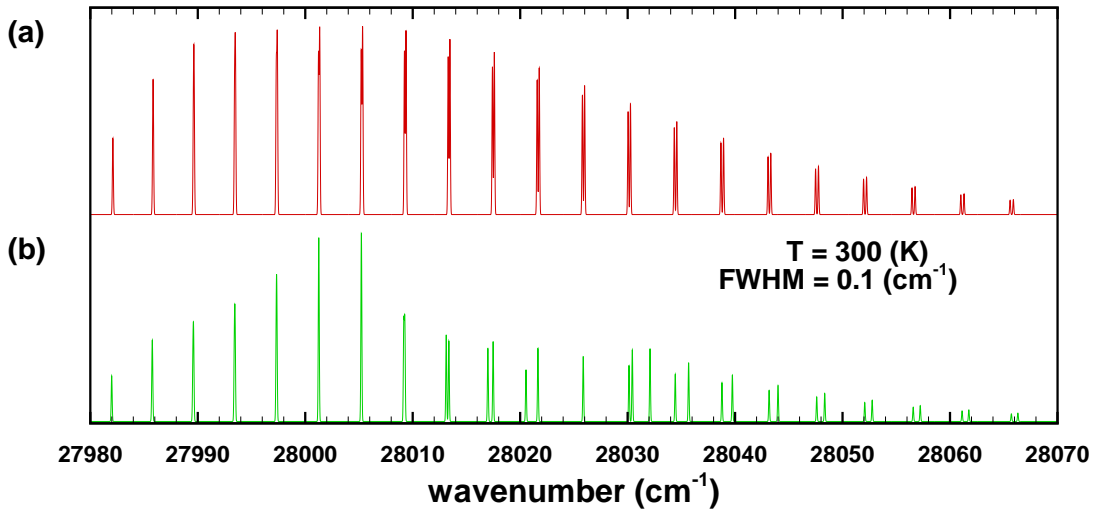}
\end{center}
\caption{Synthetic emission spectra showing the influence of inclusion of the $v=17, A \, ^2\Pi$ basis in the upper $v=5$ state of the CN violet (5,4) band. In the upper spectrum, (a), the upper states are pure $^2\Sigma^+$. The $v=17, A \, \, ^2\Pi$ energy eigenvalues lie very near the $v=5, B \, ^2\Sigma^+$ eigenvalues, and this explains the large influence of the $A \, ^2\Pi$ basis. In the lower spectrum, (b), the upper states are treated as the sum $c_{\, \Sigma} \, ^2\Sigma^+ + c_{\Pi} \, ^2\Pi$ with $c_{\, \Sigma} \gg c_{\Pi}$.  Only $R$ branch lines are shown here, including those given in Table \ref{list}.}
\label{high}
\end{figure}

\begin{figure}
\begin{center}
\includegraphics[scale=0.75]{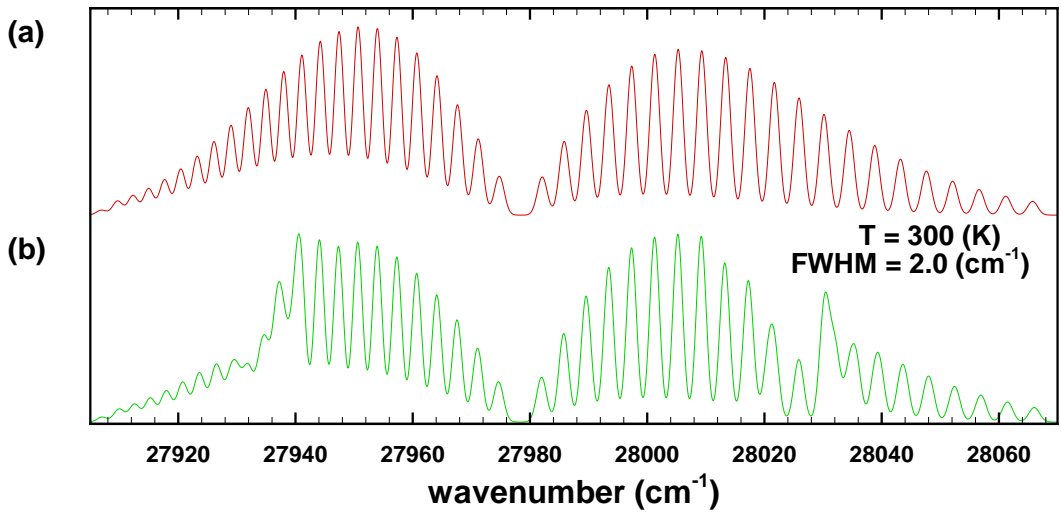}
\end{center}
\caption{The lower resolution spectra include both the $P$ and $R$ branches. (a) pure, (b) addition of a small amount of \ $^2\Pi$ to the upper basis affects the lower spectrum of the violet (5,4) band even at low resolution. }
\label{low}
\end{figure}

For the computation of rotational spectra, the square of transition moments are numerically computed using the eigenvectors of upper and lower Hamiltonians. This approach can also be selected in the new PGOPHER program \cite{Pgopher}. For the diatomic molecule, the results effectively yield the H\"onl-London factors yet we do not utilize tabulated H\"onl-London factors that are available in standard textbooks. Table \ref{list} and Figures \ref{high} and \ref{low} compare results obtained with and without taking into account the mixing. Results of modeling the angular momentum states of the upper $v=5$ vibrational level as a mixture of $^2\Sigma$ and $^2\Pi$ Hund's case (a) basis functions, a so-called ``de-perturbation'' or perturbation analysis, agree well that of Ito \textit{et al.} \cite{Ito5-4} whose used the line position measurements of Engleman \cite{Engleman74}. The 100 lines of the more recent data of Ram \textit{et al.} \cite{RDWEAB} were fitted with a standard deviation of $0.025$ cm$^{-1}$. Failure to include spin-orbit mixing of the $B \, ^2\Sigma^+$ and $A \, ^2\Pi$ basis states increased the standard deviation to $0.25$ cm$^{-1}$.

\begin{table}[h]
\center
\caption{Lines in the CN $B \, ^2\Sigma^+ - X \, ^2\Sigma^+$ (5,4) band near the perturbation. $\tilde\nu$ are the fitted line positions, $S(J',J)$ are the rotational line strengths computed in the fitting algorithm. $S^{(0)}(J',J)$ and $\Delta\tilde\nu^{(0)}$ are the line strengths and errors in the fitted line positions, respectively, when the off-diagonal spin-orbit coupling constants $\langle AL+\rangle$ and $\langle BL+\rangle$ are set equal to 0. Spin-orbit mixing of $B \, ^2\Sigma^+$ and $A \, ^2\Pi$ shifts the upper $e$ parity levels. An error in the $\tilde\nu(J',J)$ associated with these upper $e$ parity levels is produced if the mixing is ignored. A relatively large fractional error [\emph{e.g.}, -3.974/17.455 versus -1.870/28032 for $R_{11}(12.5)$] can occur in the rotational line strengths, $S(J',J)$.}
\begin{tabular}{rrrrrrrrr}
\multicolumn{9}{c}{} \\
\hline\hline
\multicolumn{1}{c}{$J'$} &
\multicolumn{1}{c}{$J$} &
\multicolumn{1}{c}{} &
\multicolumn{1}{c}{$p'$} &
\multicolumn{1}{c}{$\tilde\nu$} &
\multicolumn{1}{c}{$S_{J'J}$} &
\multicolumn{1}{c}{$\Delta\tilde\nu$} &
\multicolumn{1}{c}{$S^{(0)}_{J'J}$} &
\multicolumn{1}{c}{$\Delta\tilde\nu^{(0)}$} \\
\hline
 9.5$\ $ & 8.5 & $R_{11}$ & $-e$  & 28013.117 & 9.474 & -0.010 & 9.474 & 0.337 \\
 9.5$\ $ & 8.5 & $R_{22}$ & $+f$ & 28017.421 & 9.474 & 0.001 & 9.474 & -0.059 \\
10.5$\ $ &  9.5 & $R_{11}$ & $+e$ & 28016.992 & 9.1988 & -0.004 & 10.476 & 0.600 \\
10.5$\ $ & 9.5 & $R_{22}$ & $-f$ & 28021.651 & 11.171 & -0.000 & 10.476 & -0.067 \\
11.5$\ $ & 10.5 & $R_{11}$ & $-e$ & 28020.540 & 7.868 & -0.041 & 11.478 & 1.193 \\
11.5$\ $ & 10.5 & $R_{22}$ & $+f$ &  28025.866 & 12.240 & 0.006 & 11.478 & -0.067 \\
12.5$\ $ & 11.5 & $R_{22}$ & $-f$ &  28030.125 & 13.288 & 0.007 & 12.480 & -0.072 \\
12.5$\ $ & 11.5 & $R_{11}$ & $+e$ & 28030.431 & 13.812 &  & 12.480 \\
13.5$\ $ & 12.5 & $R_{11}$ & $-e$ & 28032.081 & 17.455 & -0.053 & 13.481 & -1.870 \\
13.5$\ $ & 12.5 & $R_{22}$ & $+f$ & 28034.428 & 14.325 & 0.011 & 13.481 & -0.073 \\
14.5$\ $ & 13.5 & $R_{11}$ & $+e$ & 28035.672 & 17.919 & -0.005 & 14.483 & -1.102 \\
14.5$\ $ & 13.5 & $R_{22}$ & $-f$ & 28038.773 & 15.356 & 0.013 & 14.483  & -0.076 \\
15.5$\ $ & 14.5 & $R_{11}$ & $-e$ & 28039.742 & 18.442 & 0.007 & 15.484 & -0.807 \\
15.5$\ $ & 14.5 & $R_{22}$ & $+f$ & 28043.161 & 16.383 & 0.009 & 15.484 & -0.084 \\
16.5$\ $ & 15.5 & $R_{11}$ & $+e$ & 28043.989 & 19.132 & 0.011 & 16.485 & -0.655 \\
16.5$\ $ & 15.5 & $R_{22}$ & $-f$ & 28047.590 & 17.405 & 0.006 & 16.485 & -0.091 \\
\hline\hline
\end{tabular}
\label{list}
\end{table}

\begin{figure}[h]
\begin{center}
\includegraphics[scale=0.75]{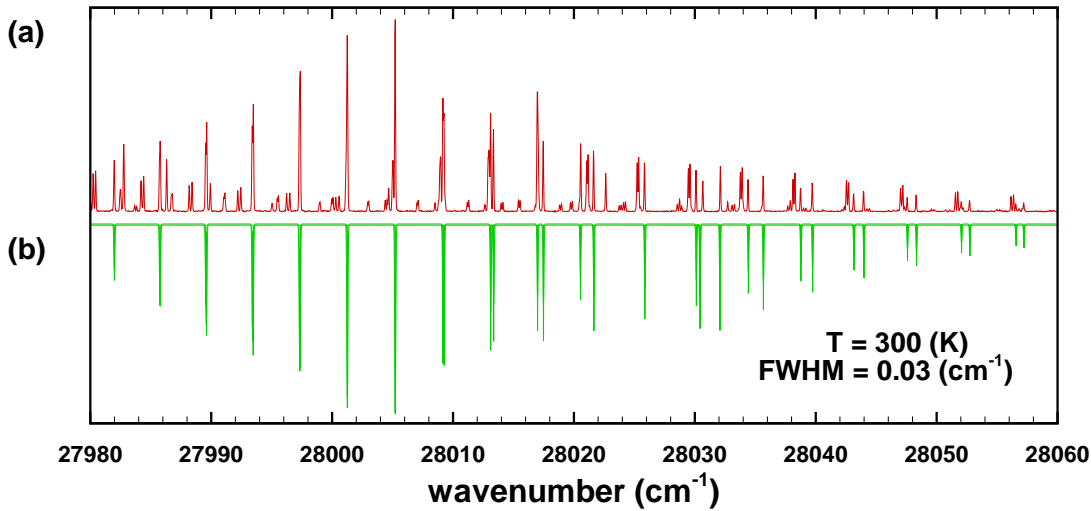}
\end{center}
\caption{Comparison of measured and simulated spectra. (a) Segment of the recorded \cite{RDWEAB} Fourier transform spectrum 920212R0.005 \cite{NSO}, (b) computed spectrum for a temperature of 300 K and a spectral resolution of 0.03 cm$^{-1}$. The computed (5.4) band is flipped vertically to show how  the predicted line positions of the R-branch match the vacuum wavenumbers of the experimental spectrum.}
\label{comparison}
\end{figure}

The table and synthetic spectra reveal that the changes caused by spin-orbit mixing are relatively very much larger for the rotational line strengths, $S(J',J)$, than for the line positions, $\tilde{\nu}$. The simulation results compare nicely with measured spectra \cite{RDWEAB} available from the National Solar Observatory (NSO) at Kitt Peak \cite{NSO}. Figure \ref{comparison} displays the recorded and simulated spectra for a resolution of 0.03 cm$^{-1}$.

The influence of $^2\Sigma^+ + \, ^2\Pi$ mixing on the rotational line strengths, $S(J',J)$, was recognized because computation of $S(J',J)$ is an integral part of the unique line position fitting algorithm. Upper and lower Hamiltonian matrices in the Hund's case (a) basis are numerically diagonalized, and the spectral line vacuum wavenumber $\tilde\nu$ is the difference between upper and lower Hamiltonian eigenvalues. To determine which of the many eigenvalue differences represent allowed spectral lines, the  factor $S(J',J)$ is computed from the upper and lower eigenvectors for each eigenvalue difference. A non-vanishing $S(J',J)$ denotes an allowed diatomic spectral line. Parity partitioned effective Hamiltonians are not used. Parity and branch designation are not required in the fitting algorithm. Input data to the fitting program is a table of vacuum wavenumber $\tilde\nu$ versus $J'$ and $J$. The non-vanishing of the rotational strength is the only selection rule used. Applications of this rule leads to the establishment of spectral data bases for diatomic molecular spectroscopy of selected transitions \cite{Parigger}. Over and above the PGOPHER program \cite{Pgopher}, there are other extensive efforts in predicting diatomic molecular spectra including for instance the so-called \textsc{Duo} program \cite{DUO} for diatomic spectroscopy.

\section{Wigner-Witmer diatomic eigenfunction}
The Hund's case (a) basis functions were derived from the Wigner and Witmer \cite{Wigner&Witmer} diatomic eigenfunction,
\begin{equation}
\langle \rho, \zeta, \chi, \mathbf{r}_2, \dots, \mathbf{r}_N, r, \theta, \phi \, | nvJM\rangle 
= \sum_{\Omega=-J}^J \langle \rho, \zeta,  \mathbf{r}'_2, \dots, \mathbf{r}'_N, r \, | nv \rangle \,  D_{M \Omega}^{J^{\scriptstyle *}} (\phi, \theta, \chi).
\label{WWeig}
\end{equation}

\noindent The coordinates are $\rho$ the distance of one electron (the electron arbitrarily labeled 1 but it could be any one of the electrons) from the internuclear vector $\mathbf{r}(r, \theta, \phi)$, the distance $\zeta$ of that electron above or below the plane perpendicular to $\mathbf{r}$ and passing through the center of mass of the two nuclei (the coordinate origin), the angle $\chi$ for rotation of that electron about the internuclear vector $\mathbf{r}$, and the remaining electronic coordinates $\mathbf{r}_2, \dots, \mathbf{r}_N$ in the fixed and $\mathbf{r}'_2, \dots, \mathbf{r}'_N$ in the rotating coordinate system. The vibrational quantum number $v$ has been extracted from the quantum numbers collection $n$ which represents all required quantum numbers except  $J$, $M$, $\Omega$, and $v$.  The Wigner-Witmer diatomic eigenfunction has no application in polyatomic theory, but for the diatomic molecule the exact separation of the Euler angles is a clear advantage over the Born-Oppenheimer approximation for the diatomic molecule in which the angle of electronic rotation, $\chi$, is unnecessarily separated from the angles describing nuclear rotation, $\theta$ and $\phi$. Equation (\ref{WWeig}) can be derived by writing the general equation for coordinate (passive) rotations $\alpha$, $\beta$, and $\gamma$ of the eigenfunction, replacing two generic coordinate vectors with the diatomic vectors $\mathbf{r}(r, \theta, \phi)$ and $\mathbf{r}'(\rho, \zeta, \chi)$, and equating the angles of coordinate rotation to the angles of physical rotation $\phi$, $\theta$, and $\phi$. The general equation for coordinate rotation holds in isotropic space, and therefore the quantum numbers $J$, $M$, and $\Omega$ in the Wigner-Witmer eigenfunction include all electronic and nuclear spins. If nuclear spin were to be included, $J$, $M$, and $\Omega$ would be replaced by $F$, $M_F$, and $\Omega_F$, but hyperfine structure is not resolved in the $(5,4)$ band data reported by \cite{RDWEAB}, and Eq.~(\ref{WWeig}) is written with the appropriate spectroscopic quantum numbers.

It is worth noting that the rotation matrix element $D_{M \Omega}^J (\phi, \theta, \chi)$ and its complex conjugate $D_{M \Omega}^{J^{\scriptstyle *}} (\phi, \theta, \chi)$ do not fully possess the mathematical properties of quantum mechanical angular momentum. It is well known that a sum of Wigner $D$-functions is required to build an angular momentum state. The equation
\begin{equation}
J'_{\pm} \, D_{M \Omega}^{J^{\scriptstyle *}} (\phi, \theta, \chi)
= \sqrt{J(J+1) - \Omega(\Omega \mp 1)} \, D_{M, \Omega \mp 1}^{J^{\scriptstyle *}} (\phi, \theta, \chi)
\label{useful}
\end{equation}

\noindent is not a phase convention \cite{Zare1973} \cite{Brown&Howard1976} \cite{Lefebvre-Brion&Field} but a mathematical result readily obtained from Eq.~(\ref{WWeig}) and
\begin{equation}
J'_{\pm} \, |J\Omega\rangle = \sqrt{J(J+1) - \Omega(\Omega \pm 1)} \, |J, \Omega\pm 1 \rangle,
\label{Jpm}
\end{equation}

\noindent in which the prime on the operator $J'_{\pm}$ indicates that it is written in the rotated coordinate system where the appropriate magnetic quantum number $\Omega$.

\section{Hund's basis functions}
The Hund's case (a) basis function based upon the Wigner-Witmer diatomic eigenfunction is
\begin{equation}
|a\rangle = \langle \rho,\zeta, \chi, \mathbf{r}'_2, \dots, \mathbf{r}'_{\mathcal{N}}, r, \theta, \phi \, | nvJMS\Lambda\Sigma\Omega\rangle 
= \sqrt{\frac{2J+1}{8 \pi^2}} \langle \rho, \zeta, \mathbf{r}'_2, \dots, \mathbf{r}'_{\mathcal{N}}, r \, | nv\rangle \, |S\Sigma\rangle \, D_{M \Omega}^{J^{\scriptstyle*}} (\phi, \theta,\chi).
\label{Casea}
\end{equation}

\noindent As noted above, a sum of $|a\rangle$ basis functions is required to build an eigenstate of angular momentum. The basis function would also not be an eigenstate of the parity operator. The case (a) matrix elements, $p^{(a)}_{ij}$, of the parity operator $\mathcal{P}$,
\begin{equation}
p^{(a)}_{ij} = p_{\, \Sigma} (-)^J \delta(J_i J_j) \, \delta(\Omega_i, -\Omega_j) \, \delta(\Lambda_i, -\Lambda_i)  \, \delta(n_i n_j),
\label{aParity}
\end{equation}

\noindent show that a single $|a\rangle$ basis function is not an eigenstate of parity. The procedure called parity symmetrization adds $|JM\Omega\rangle$ and $|JM,-\Omega\rangle$ basis functions thereby destroying the second magnetic quantum number $\Omega$ and yielding a function which at least possesses the minimal mathematical properties of an eigenstate of angular momentum, parity, and the other members of the complete set of commuting operators. The general procedure would be to continue adding basis functions to the upper and lower bases until eigenvalue differences between the upper and lower Hamiltonians accurately predict measured line positions.

\section{The upper Hamiltonian matrix for the (5,4) band}
Electronic spin $\mathbf{S}$ interactions with electronic orbital momentum $\mathbf{L}$ and nuclear orbital momentum $\mathbf{R}$ produce both diagonal and off-diagonal matrix elements in the Hund's case (a) representation of the Hamiltonian. The off-diagonal elements connect different basis states. For example, both of the mentioned spin orbit interactions connect $^2\Sigma^+$ and $^2\Pi$. Because van Vleck transformed Hamiltonians are not used, the appropriate parameters for the strength of these interactions are $\langle AL+\rangle$ and $\langle BL+\rangle$. Table \ref{pars} lists the molecular parameters used in the Hamiltonian matrices.
Tables \ref{NoSpinOrbit} and \ref{SpinOrbit} show the Hamiltonian matrices without
and with spin-orbit interactions, respectively.

\begin{table}
\center
\begin{tabular}{r|rrr}
\hline\hline
 & \multicolumn{1}{c}{$X^2\Sigma^+$} & \multicolumn{1}{c}{$B^2\Sigma^+$}        & \multicolumn{1}{c}{$A^2\Pi$} \\
 & \multicolumn{1}{c}{$v=4$} & \multicolumn{1}{c}{$v=5$} & \multicolumn{1}{c}{$v=17$} \\
\hline
$B_v$ & 1.820866(13) & 1.845727(13) & 1.404833 \\
$D_v$ & $6.172(36) \times 10^{-6}$ & $8.003(38) \times 10^{-6}$ & $5.66 \times 10^{-6}$ \\
$A_v$ &  &  & $-50.5253$ \\
$\gamma_v$ & $-1.98(43) \times 10^{-4}$ & $-1.921(44) \times 10^{-2}$ \\
$\gamma_{D_v}$ & $-1.98(43) \times 10^{-4}$ \\
$T_v$ & $8011.7871$ & $35990.1780(25)$ & $36010.5732$ \\
$<AL+>$ & & $4.25(0.03)$ \\
$<BL+>$ & & $0.0205(0.001)$ \\
\hline\hline
\end{tabular}
\caption{Molecular parameters used in this work that relies on Hamiltonians that are not parity-partitioned. Values not followed by a number in parenthesis were held fixed or an error estimate was not computed. A value in parenthesis is the standard deviation in the fitted value. Parameters for the $A^{\, 2}\Pi$ state were fitted by the Nelder-Mead minimization algorithm using values given by Brooke \textit{et al.} \cite{JSABrookeCN} as trial values. Error estimates were not computed, and the values of Brooke \textit{et al.} \cite{JSABrookeCN} were only very slightly changed.}
\label{pars}
\end{table}

\begin{table}
\begin{center}
\begin{tabular}{rrrr|rrrrrr}
\hline \hline
& & & $v$ & 5 & 5 & 17 & 17 & 17 & 17 \\
& & & $\Lambda$ & 0 & 0 & -1 & -1 & 1 & 1 \\
& & & $\Sigma$ & -0.5 & 0.5 & -0.5 & 0.5 & -0.5 & 0.5 \\
$v$ & $\Lambda$ & $\Sigma$ & $\Omega$ & -0.5 & 0.5 & -1.5 & -0.5 & 0.5 & 1.5 \\
\hline
5 & 0 & -0.5 & -0.5 & 36351.6409 & -25.6707 & 0 & 0 & 0 & 0 \\
5 & 0 & 0.5 & 0.5 & -25.6707 & 36351.6409 & 0 & 0 & 0 & 0 \\
17 & -1 & -0.5 & -1.5 & 0 & 0 & 36257.6340 & -19.5866 & 0 & 0 \\
17 & -1 & 0.5 & -0.5 & 0 & 0 & -19.5866 & 36310.9646 & 0 & 0 \\
17 & 1 & -0.5 & 0.5 & 0 & 0 & 0 & 0 & 36310.9646 & -19.5866 \\
17 & 1 & 0.5 & 1.5 & 0 & 0 & 0 & 0 & -19.5866 & 36257.6340 \\
\hline
& & \multicolumn{2}{r|}{$E_{nvJ}$}&36377.3116&36325.9702&36251.2135&36317.3851&36317.3851&36251.2135 \\
\hline \hline
\end{tabular}
\caption{Hamiltonian matrix for states modeled as the mixture of \, $^2\Sigma^+$ and $^2\Pi$ basis states. Off-diagonal spin-orbit coupling has been removed.  Consequently, the $ 2 \times 2$  matrices along the main
diagonal are independent, and could be individually diagonalized. The bottom row contains the energy eigenvalues.
Using matrices like these to model upper states of the CN violet (5,4) band, the 100 experimental spectral lines reported by Ram \textit{et al.} \cite{RDWEAB} were fitted with a standard deviation of 0.25 ${\rm cm}^{\, -1}$. This Hamiltonian was computed for $\langle AL+\rangle = \langle BL+\rangle = 0$ but is otherwise identical to the Hamiltonian in Table \ref{SpinOrbit}. Standard Hund's case (a) matrix elements \cite{Zare1973} \cite{Lefebvre-Brion&Field} were used.}
\label{NoSpinOrbit}
\end{center}
\end{table}

\begin{table}
\begin{center}
\begin{tabular}{rrrr|rrrrrr}
\hline \hline
& & & $v$ & 5 & 5 & 17 & 17 & 17 & 17 \\
& & & $\Lambda$ & 0 & 0 & -1 & -1 & 1 & 1 \\
& & & $\Sigma$ & -0.5 & 0.5 & -0.5 & 0.5 & -0.5 & 0.5 \\
$v$ & $\Lambda$ & $\Sigma$ & $\Omega$ & -0.5 & 0.5 & -1.5 & -0.5 & 0.5 & 1.5 \\
\hline
5 & 0 & -0.5 & -0.5 & 36351.6409 & -25.6707 & 2.8566 & 2.3274 & 2.8639 & 0 \\
5 & 0 & 0.5 & 0.5 & -25.6707 & 36351.6409 & 0 & 2.8639 & 2.3274 & 2.8566 \\
17 & -1 & -0.5 & -1.5 & 2.8566 & 0 & 36257.6340 & -19.5866 & 0 & 0 \\
17 & -1 & 0.5 & -0.5 & 2.3274 & 2.8639 & -19.5866 & 36310.9646 & 0 & 0 \\
17 & 1 & -0.5 & 0.5 & 2.8639 & 2.3274 & 0 & 0 & 36310.9646 & -19.5866 \\
17 & 1 & 0.5 & 1.5 & 0 & 2.8566 & 0 & 0 & -19.5866 & 36257.6340 \\
\hline
& & \multicolumn{2}{r|}{$E_{nvJ}$}&36377.3957&36327.7869&36250.9625&36317.3525&36315.8194&36251.1620 \\
\hline \hline
\end{tabular}
\caption{Off-diagonal spin-orbit coupling $6 \times 6$ matrix. From the three independent $2 \times 2$ matrices of Table \ref{NoSpinOrbit}, the off-diagonal matrix elements mix the Hund's case (a) basis states, and the standard deviation of the spectral line fit mentioned in Table \ref{NoSpinOrbit} is reduced by a factor of 10 to 0.025 ${\rm cm}^{\, -1}$. The spin-orbit coupling constants $\langle AL+\rangle = 4.25 (0.03)$ and $\langle BL+\rangle = 0.205 (0.001)$ were used in computation of this Hamiltonian. This single $6 \times 6$ matrix describing $^2\Pi - ^2\Sigma^+$ mixing can be compared with the two $3 \times 3$ parity partitioned matrices of Brown and Carrington \cite{Brown&Carrington}.}
\label{SpinOrbit}
\end{center}
\end{table}

\section{A diatomic line position fitting algorithm}
A basic tool for the diatomic spectroscopist is a computer program that accepts a table of experimentally measured vacuum wave numbers $\tilde\nu_{\rm exp}$ versus $J'$ and $J$, and outputs a set of molecular parameters with which one can reproduce the $\tilde\nu_{\rm exp}$ with a standard deviation comparable to the estimated experimental error. In practice, an experimental line list frequently shows gaps, \textit{viz.} spectral lines are missing. Following a successful fitting process, one can use the molecular parameters to predict all lines. A computed line list is especially useful when it includes the Condon and Shortley \cite{Condon&Shortley} line strength from which the Einstein coefficients and oscillator strength \cite{Hilborn} \cite{Thorne} and the HITRAN line strength \cite{Rothman1996} can be calculated. A feature of the line fitting program described below is its use of non-zero rotational strengths (see Eq.~(\ref{HLF}) below) to mark which of the many computed differences between upper and lower term values represents the vacuum wavenumber of an allowed spectral line. Consequently, the fitting process creates a complete line list including rotational factors.  Parity plays no part in the fitting process, but the same orthogonal matrix that diagonalizes the case (a) Hamiltonian matrix will also diagonalize the case (a) parity matrix whose elements are given in Equation (\ref{aParity}). The $p = \pm 1$ parity eigenvalue becomes a computed quantity, and the $e/f$ parity designation is established from the parity eigenvalue using the accepted convention Brown \textit{et al.} \cite{Brown_ef1975}.

Trial values of upper and lower state molecular parameters, typically taken from previous works by other for the band system in question, are used to compute upper $H$' and lower $H$  Hamiltonian matrices in the case (a) basis given by Eq.~(\ref{Casea}) for specific values of $J'$ and $J$.  The upper and lower Hamiltonians are numerically diagonalized,
\begin{subequations}
\begin{align}
T' &= \tilde{U'} \, H' \, U' \\
T &= \tilde{U} \, H \, U
\end{align}
\end{subequations}

\noindent giving the upper $T'$ and lower $T$ term values.  The vacuum wavenumber $\tilde\nu$ is determined,
\begin{equation}
\tilde\nu_{ij} = T'_i - T_j,
\end{equation}

\noindent and the rotational strength is evaluated,
\begin{equation}
S_{ij}(J',J) = (2J+1) \left| \sum_n \sum_m \tilde{U'}_{in} \langle J \Omega; q, \Omega'-\Omega \, | J' \Omega'\rangle \, U_{mj} \, \delta(\Sigma'_n \Sigma_m) \right|^2.
\label{HLF}
\end{equation}

\noindent The degree of the tensor operator, $q$, responsible for the transitions amounts to $q = 1$ for electric dipole transitions. For a non-zero rotational factors, $S(J',J)$, the vacuum wavenumber $\tilde\nu_{ij}$ is added to a table of computed line positions to be compared with the experimental list $\tilde\nu_{\rm exp}$ versus $J'$ and $J$. The Clebsch-Gordan coefficient, $\langle J \Omega; q, \Omega'-\Omega \, | J' \Omega'\rangle$, is the same one appearing in the pure case (a) - case (a) formulae for $S(J',J)$. For a specific values of $J'$ and $J$, one constructs tables for $\tilde\nu_{\rm exp}$ and computed $\tilde\nu_{ij}$. The errors $\Delta\tilde\nu_{ij}$,
\begin{equation}
\Delta\tilde\nu_{ij} = \tilde\nu_{ij} - \tilde\nu_{\rm exp},
\end{equation}

\noindent are computed where each $\tilde\nu_{ij}$ is the one that most closely equals one of the $\tilde\nu_{\rm exp}$. Once values of $\tilde\nu_{ij}$ and $\tilde\nu_{\rm exp}$ are matched, each is marked unavailable until a new list of $\tilde\nu_{ij}$ is computed. The indicated computations are performed for all values of $J'$ and $J$ in the experimental line list, and corrections to the trial values of the molecular parameters are subsequently determined from the resulting $\Delta\tilde\nu_{ij}$. The entire process is iterated until the parameter corrections become negligibly small. As this fitting process successfully concludes, one obtains a set of molecular parameters that predict the measured line positions $\tilde\nu_{\rm exp}$ with a standard deviations that essentially equal the experimental estimates for the accuracy of the $\tilde\nu_{\rm exp}$.

\section{Discussion}
The influence on intensities in the (5,4) band of the CN violet system caused by the weak spin-orbit mixing, Figs.~\ref{high} and \ref{low}, is significantly larger than initially anticipated. This was noticed because computation of the
rotational strengths is an integral part of our line position fitting program. The eigenvectors that diagonalize the Hamiltonian to yield fitted line
position $\tilde\nu$ also yield $S(J',J)$.
In established diatomic molecular practice, H\"onl-London factors are determined independently of line positions. Analytical approximations utilize the parameter $Y = A / B$ to account for the influence of spin-orbit interaction on $S(J',J)$. Kov{\'a}cs \cite{Kovacs} gives many examples, Li \textit{et al.} \cite{GangLI} give a more recent application. These analytical approximations can accurately account for intermediate spin-orbit coupling which smooth transitions between case (a) and case (b) with increasing $J'$ and $J$, but show limited sensitivity to abrupt changes in $S(J',J)$ near perturbations such as those seen the $(5,4)$ band in the CN violet system.

\section{Conclusion}
The Wigner-Witmer diatomic eigenfunction makes it possible to form an exact, mathematical connection between computation of $\tilde\nu$ and $S(J',J)$ in a single algorithm. The concept of the non-vanishing rotational strengths as the omnipotent selection rule initially conceived  as a simplifying convenience in a computer algorithm is now seen to be more valuable, as evidenced in this work's analysis of the CN (5,4) band perturbations by isolating a specific branch. Future work is planned for comparisons of the CN (10,10) band spectra that include perturbation and that show promising agreements with experiments and PGOPHER predictions.


\section{Acknowledgments}
One of us (CGP) acknowledges support in part by UTSI's Accomplished Center of Excellence in Laser Applications.

\section*{References}

\bibliographystyle{elsarticle-num}

\end{document}